\documentclass[aps,preprint]{revtex4-1}

\usepackage{epsfig,amsmath}
\usepackage{subfigure}
\usepackage{graphicx}
\usepackage{dcolumn}
\usepackage{stmaryrd}
\usepackage{mathrsfs}
\usepackage{pifont}
\usepackage{amsthm}
\usepackage{amssymb}
\usepackage{bm}
\usepackage{latexsym}
\usepackage[colorlinks=true,linkcolor=blue,citecolor=blue]{hyperref}
\usepackage{color}
\usepackage{epstopdf}

\theoremstyle{plain}

\newcommand{\la}{\langle}
\newcommand{\ra}{\rangle}

\newcommand{\ti}{\tilde}
\newcommand{\ga}{\gamma}

\newcommand{\ka}{\kappa}
\newcommand{\da}{\dagger}
\newcommand{\De}{\Delta}

\newcommand{\om}{\omega}
\newcommand{\Om}{\Omega}
\newcommand{\de}{\delta}

\newcommand{\pa}{\partial}

\newcommand{\mau}{\mathcal{U}}

\def\jpa#1{{ J.\ Phys.\ A} {\bf#1}}
\def\pra#1{{ Phys.\ Rev. A\/} {\bf#1}}
\def\prb#1{{ Phys.\ Rev. B\/} {\bf#1}}

\def\prl#1{{ Phys.\ Rev.\ Lett.} {\bf#1}}

\def\pla#1{{ Phys.\ Lett. A\/} {\bf#1}}
\def\rmp#1{{ Rev. \ Mod. \ Phys.} {\bf#1}}
\def\nat#1{{ Nature} {\bf#1}}

\begin{document}

\title{One-component quantum mechanics and dynamical leakage-free paths}

\author{Jun Jing}
\affiliation{Department of Physics, Zhejiang University, Hangzhou 310027, Zhejiang, China}

\author{Lian-Ao Wu}
\email{Email address: lianao.wu@ehu.es}
\affiliation{Department of Physics, The Basque Country University (EHU/UPV), PO Box 644, 48080 Bilbao, Spain and Ikerbasque, Basque Foundation for Science, 48011 Bilbao, Spain}

\date{\today}

\begin{abstract}
We derive an exact one-component equation of motion for the probability amplitude of a target time-dependent state, and use the equation to reformulate quantum dynamics and control for both closed and open systems. Using the one-component equation, we show that an unexpected time-dependent leakage-free path can be induced and we capture the essential quantity in determining the effect of decoherence suppression. Our control protocol based on the nonperturbative leakage elimination operator provides a unified perspective connecting some subtle, popular and important concepts of quantum control, such as dynamical decoupling, quantum Zeno effect, and adiabatic passage. The resultant one-component equation will promise significant advantages in both quantum mechanics and control.
\end{abstract}

\maketitle

\section{Introduction}

Quantum mechanics is based on postulates, such as (a) quantum dynamics governed by the Schr\"odinger equation, (b) quantum measurement~\cite{WisemanBook} by projection or postselection, and (c) boundary conditions of the system wave-function or density matrix that induce interesting phenomenon such as topological insulator~\cite{LaughlinPRB81}. Most of the existing quantum control protocols~\cite{BrumerBook} involve with the first two postulates. Dynamical control~\cite{AgarwalPRL01,AgarwalPRA01} that is usually realized by laser pulse sequence, includes, such as bang-bang (BB) control~\cite{ViolaPRL99,KurizkiPRL04,UhrigPRL09,LidarPRL10}, when the evolution operator of the system and environment is intersected by unitary and instantaneous control operations; the nonperturbative dynamical decoupling~\cite{JingPRA13,JingSR13,JingSR14,JWPRL15}, when the system Hamiltonian is modified by the time-dependent pulse sequence; and the adiabatic passage~\cite{BornZPhys28,WuzhaoyanPRA05,ShapiroRMP07} based on the slowly-varying Hamiltonian and the avoiding of the level crossings. Nonunitary quantum control relies on the projection methods, such as quantum Zeno-like effect~\cite{SudarshanJMP77,KurizkiNAT00,BrionPRA05}, when the total evolution operator of the system and environment is frequently interrupted by an operation projecting the system to the desired state or subspace. These methods seem to be dramatically different from each other.

In this paper, we show that the aforementioned control protocols and the corresponding dynamics of the interested system can be uniformly formulated by an exact one-component equation that properly traces the probability amplitude of the target time-dependent state $|A(t)\ra$ or $|A(t)\ra\ra$, which is a vector living in a normal Hilbert space or a superoperator space~\cite{JingPRA14,LidarPRA05}. Using this equation, we can obtain a sufficient condition for creating a {\em leakage-free path} (LFP), that conceptually generalizes the conventional decoherence-free subspace~\cite{KnightPRA97,LidarBook03}. Also, this condition extends the existing conditions for realizing dynamical decoupling.

The paper is outlined as follows. In Sec.~\ref{onecomponent}, we present the theoretical framework of one-component quantum mechanics via the one-variable dynamical equation based on the P-Q partition. In Sec.~\ref{LDP}, we propose a universal control protocol towards LFP. We show in Sec.~\ref{diss} that the conventional bang-bang control, quantum Zeno effect, and the adiabatic passage could be united under the condition for LFP. In Sec.~\ref{app}, our control protocol is applied to various pedagogical models, including the two-level system, the spin-spin-bath model, the multiple level system under non-Markovian environment. And we conclude our work in Sec.~\ref{conc}.

\section{One-component quantum mechanics}\label{onecomponent}

We consider a generic {\em linear} equation of motion, $\pa_{t}\mathcal{X}=\mathcal{MX}$, exemplified by matrix representations of the Schr\"odinger equation, the stochastic Schr\"odinger equation that combines the Schr\"odinger equation of the entire system and quantum measurements on its environment or bath, such as quantum-state-diffusion equation~\cite{GardinerBook,DiosiPLA97,DiosiPRA98,StrunzPRL99}, the perturbative master equation~\cite{BreuerBook}, and even a classical equation for linear-interacting harmonic oscillators. We derive the one-component equation via the Feshbach P-Q partitioning technique~\cite{WuPRL09,Wuonline,JingPRA12}. The $n$-dimensional vector $\mathcal{X}$ and the $n \times n$ dynamic matrix $\mathcal{M}$ can be partitioned as
\begin{equation}\label{Mt}
\mathcal{X}=\left[\begin{array}{c} P \\ \hline Q
\end{array}\right], \quad
\mathcal{M}=\left(\begin{array}{c|c}
      h & R \\ \hline
      W & D
    \end{array}\right),
\end{equation}
where $P$ is the probability amplitude of the one-dimensional target state $|A(t)\ra$, and the $(n-1)$-dimensional $Q$ resides in the subspace orthogonal to $|A(t)\ra$.

\begin{table}[htbp]\centering
\begin{tabular}{|c|c|c|} \hline
   & $\mathcal{M}$ & $\mathcal{X}$   \\ \hline
Classical harmonic oscillator & $\left[\begin{array}{cc}0 & 1/m \\  -m\om^2 & 0 \end{array}\right]$ & $\left(\begin{array}{c}q \\ p \end{array}\right)$ \\ \hline
Closed quantum system & $-iH$ &  $|\psi\ra$ \\ \hline
Open quantum system I & $\mathcal{L}$ & $|\rho\ra\ra$  \\ \hline
Open quantum system II & $H_{\rm eff}$ & $|\psi(z^*)\ra$  \\ \hline
\end{tabular}
\caption{A list of examples of linear dynamical equations in physics. (1) Hamiltonian mechanics for classical harmonic oscillator model, where $p$ and $q$ are generalized coordinates and momentum. (2) Schr\"odinger equation for closed quantum system. (3) Liouville equation for open quantum system~\cite{Lindblad}, where $\mathcal{L}$ is a Liouvillian super-operator, where $\rho$ is the density matrix of the system. For a $d$-dimensional system, $|\rho\ra\ra\equiv(\rho_1, \rho_2, \cdots, \rho_{d^2})'$. In this case, the dimensionality of $\mathcal{X}$ is $n=d^2$ and $|A(t)\ra\ra\equiv P(t)|A(t)\ra\la A(t)|$ in normal representation. (4) Stochastic Schr\"odinger equation for open quantum system, e.g., the quantum-state-diffusion equation, where $H_{\rm eff}$ is the non-Hermitian effective Hamiltonian and $|\psi(z^*)\ra\equiv\la z|\Psi\ra$ is a state of system obtained by the inner product of a stochastic environment state $|z\ra$ and the whole system state $|\Psi\ra$~\cite{DiosiPLA97,DiosiPRA98,StrunzPRL99}. }\label{leq}
\end{table}

Table~\ref{leq} shows various physical realizations of the linear equation of motion. Suppose that initially $P(0)=1$ and $Q(0)=0$. Equation~(\ref{Mt}) could be decomposed into
\begin{equation}
\pa_tP=hP+RQ, \quad \pa_tQ=WP+DQ.
\end{equation}
The formal solution is
\begin{equation}
\pa_tP(t)=h(t)P(t)+\int_0^tdsg(t,s)P(s),
\end{equation}
where $g(t,s)=R(t)G(t,s)W(s)$ with a propagator $G(t,s)=\mathcal{T}_\leftarrow\{\exp[\int_s^tds'D(s')]\}$. Alternatively, we can set $P(t)=p(t)e^{\int_0^tdsh(s)}$ and obtain a more compact formal solution
\begin{equation}\label{gts}
\pa_tp(t)=e^{iC(t)}\int_0^tdse^{-iC(s)}g(t,s)p(s)=\int_0^tdsg'(t,s)p(s),
\end{equation}
where $C(t)\equiv i\int_0^tdsh(s)$ and $g'(t,s)\equiv\exp[\int_s^th(s')ds']g(t,s)$. This one-component integro-differential equation is the core result of this work and will serve as a powerful tool in studying both dynamics and control of a given target state. If $g'(t,s)=-k^2$ with $k$ a constant number, then we recover the equation of motion for a harmonic oscillator, i.e., $\ddot{p}+k^2p=0$. And $p(t)=\cos(kt)$ when $p(0)=1$. If $g'(t,s)=-2\lambda\delta(t-s)$, then we find the ideal Markovian dynamics with $\dot{p}(t)=-\lambda p(t)$. And then $p(t)=e^{-\lambda t}p(0)$, implying one can not keep the system on the target state.

To {\em walk} on a desired target path $|A(t)\ra$, we need to eliminate the leakage between $P$ and $Q$ parts, which is equivalently to have a vanishing integral
\begin{equation}
\int_0^tdse^{-iC(s)}g(t,s)p(s)=0.
\end{equation}
In the case of quantum mechanics, $h(t)$ is a purely imaginary number so that $C(t)$ is a real function and then either real or imaginary part of $e^{-iC(s)}$ is a time-dependent function oscillating between $-1$ to $1$. In what follows, we will focus on quantum mechanical system and show that the vanishing of the integral over $e^{-iC(s)}g(t,s)p(s)$ leads to a time-dependent LFP. A trivial LFP emerges in a special case when $g(t,s)=0$. Note that $g(t,s)=0$ might hold even if none of $R(t)$, $G(t,s)$, and $W(s)$ is vanishing. In general, LFP is realized when $h(t)$ can be so manipulated that a rapid-oscillating $e^{-iC(s)}$ could cancel the effect raised by smoother functions of both $g(t,s)$ and $p(s)$ through
\begin{equation}
\int_0^tdse^{-iC(s)}g(t,s)p(s)\approx\sum_{k=0}^n(-1)^kg(t,k\tau)p(k\tau)\tau\rightarrow0,
\end{equation}
where $\tau=t/n$ with $n\rightarrow\infty$. Note here {\em smooth} means $g(t,k\tau)\approx g(t,(k+1)\tau)$ and $p(k\tau)\approx p((k+1)\tau)$. This result is supported by the Riemann$-$Lebesgue lemma~\cite{JWPRL15} provided that the characteristic frequency for the function $e^{-iC(s)}$ is larger than the cut-off frequency of $g(t,s)p(s)$.

\section{A universal quantum control protocol}\label{LDP}

Consider an either closed or open quantum system $\mathcal{S}$, whose Hilbert space is spanned by a set of time-independent or time-dependent bases $|\phi_m\ra$ ($m=0,1,2,\cdots,n-1$) and $|A(0)\ra=|\phi_0\ra$. Now we aim at control of the system to evolve along a desired quantum path characterized by a unitary transformation $\mau(t)$ in the system space, i.e., holding the system on the path $|A(t)\ra=\mau(t)|\phi_0\ra$. Thus the total Hamiltonian $H_{\rm tot}$ can be expressed in terms of the time-independent or time-dependent basis states $|\ti{\phi}_m\ra=\mau(t)|\phi_m\ra$, where $|\ti{\phi}_0\ra=|A(t)\ra$. Under the Schr\"odinger equation, $\pa_t|\psi\ra=-iH_{\rm tot}|\psi\ra$, we have
\begin{equation}
h(t)=-i\la\ti{\phi}_0|\ti{H}_{\rm tot}|\ti{\phi}_0\ra\equiv-i\la\ti{\phi}_0|[\mau(t)H_{\rm tot}\mau^\da(t)+i\dot{\mau}(t)\mau^\dagger(t)]|\ti{\phi}_0\ra,
\end{equation}
in terms of P-Q partitioning given in Eq.~(\ref{Mt}). Note $\mau(t)$ is irrelevant to $H_{\rm tot}$. For example in the scenario of open quantum system, the whole system could be decomposed into the system part that we are interested and the remainder part that are called environment. In the rotating frame with respect to the environment Hamiltonian (assumed to be time-independent), i.e., $\mau(t)=\exp(-iH_{\rm env}t)$, we always have
\begin{equation}\label{Htot}
\ti{H}_{\rm tot}=H_{\rm sys}(t)+\sum_kS_k(t)B_k(t),
\end{equation}
where $S_k(t)$ and $B_k(t)$ are Hermition operators in the space of system and environment, respectively. It turns out immediately that if $\la\ti{\phi}_0|S_k(t)|\ti{\phi}_0\ra=0$ holds for each $k$, then $h(t)=-i\la\ti{\phi}_0|H_{\rm sys}(t)|\ti{\phi}_0\ra$, which is irrespective to the operators and the size of environment. The well-known instance is the conventional leakage-free subspace for collective dephasing and dissipation~\cite{DFSLidar}. It implies therefore that {\em $h(t)$ could be under a full control without unpractically invoking control over the environment and the remainder of the system space.}

Under the bases $\{|\phi_n(t)\rangle\}$, the {\em rotating representation} Hamiltonian can always be partitioned into a similar form as $\mathcal{M}$ in Eq.~(\ref{Mt}). Suppose $\ti{H}_{\rm tot}=H_d+L$, where the block-diagonal part $H_d=h\oplus D$ and $L$ is the block-off-diagonal part consisted by $R$ and $W$. To maintain the system in the state of $|A(t)\ra$, i.e., to create a rapid-oscillating exponential function $e^{-iC(t)}$ by manipulating $h(t)$, a leakage elimination operation~(LEO) in rotating framework or a rotating LEO~\cite{WML}
\begin{equation}\label{rotLEO}
\ti{R}_L=c(t)\left[|\ti{\phi}_0\ra\la\ti{\phi}_0|-\sum_{n>0}|\ti{\phi}_n\ra\la\ti{\phi}_n|\right]
=c(t)\left[2|\ti{\phi}_0\ra\la\ti{\phi}_0|-\mathcal{I}\right]
\end{equation}
has been introduced to cancel the off-diagonal term (in charge of leakage) $L$ in the time evolution. Here $c(t)$ is the control function, which can be absorbed into $h(t)$ when Eq.~(\ref{gts}) is considered in the control protocol. It is clear to see $\{\ti{R}_L, L\}=0$ and $[\ti{R}_L, H_d]=0$, so that the LEO serves to parity-kick out the leakage $L$. Ideally, one can easily prove that if $c(t)\propto\delta(t-n\tau)$ at given times $n\tau$ $(n=1,2,\cdots)$, then
\begin{equation}
e^{-i\ti{H}_{\rm tot}[(n+1)\tau]\tau}\ti{R}^\da_Le^{-i\ti{H}_{\rm tot}(n\tau)\tau}\ti{R}_L\approx e^{-i[H_d((n+1)\tau)+H_d(n\tau)]\tau}
\end{equation}
when $\tau\rightarrow 0$ and $t\approx n\tau$. Nonperturbatively, it is shown that the nonideal pulse $c(t)$~\cite{JWPRL15} does also allow to achieve the same result. Furthermore, the general leakages, such as $\sum_kS_k(t)B_k(t)$ in the open quantum system, can be eliminated by $\ti{R}_L$.

\section{Discussion}\label{diss}

\subsection{Bang-bang control and quantum Zeno effect}

Many existing protocols targeting on decoherence-suppression or leakage elimination are found to be subsets of the control framework we proposed through the one-dimensional dynamical equation~(\ref{gts}). Moreover, our control strategy is state-independent.

That could be interpreted using a model of a two-level system subjected to unwanted disturbances from the uncontrollable Hilbert space orthogonal to that of the $P$-part in Eq.~(\ref{Mt}). To cancel the effect from the leakage Hamiltonian $L=\hat{X}(t)B_x(t)+\hat{Y}(t)B_y(t)$, where $\hat{X}(t)$ and $\hat{Y}(t)$ can flip the desired state $|A(t)\ra$ into an orthogonal state, e.g., $|A^{\perp}(t)\ra$, and $B_{x,y}(t)$ is an arbitrary environmental operator. By Trotter formula and using Eq.~(\ref{Htot}), in a short time interval $\de_1$, the system approximately evolves into $|A(\de_1)\ra-i\de_1|A^{\perp}(\de_1)\ra$, where $|A(\de_1)\ra=\mathcal{U}(\de_1)|A(0)\ra=e^{-iH_{\rm sys}(0)\de_1}|A(0)\ra$. In order to stabilize the passage of the system along the desired path, one has to insert a BB control pulse indicated by $\hat{Z}(t)$ to the system evolution after the period of free-evolution $\de_1$. Similar to $\hat{X}(t)$ and $\hat{Y}(t)$ given before, $\hat{Z}(t)$ is not necessary the Pauli matrix along the $z$-direction. It is merely required that these three operators constitute a set of generators of $SU(2)$. For instance, $\hat{Z}(t)$ can be chosen as $|A(\de_1)\ra\la A(\de_1)|-|A^{\perp}(\de_1)\ra\la A^{\perp}(\de_1)|$. To the first-order perturbation, the state of the system is now approximated as $|A(\de_1)\ra$ by virtue of $\{L, \hat{Z}(t)\}=0$. This process could be repeated many times until to the desired moment $t=\sum_j\de_j$. BB control always works as long as each interval $\de_j$ is short enough. However, it amounts to taking the zero-order perturbation for an effective Hamiltonian $H_{\rm eff}=c\hat{Z}+\ti{H}_{\rm tot}$, where $c\de=\pi/2$ and the system-bath Hamiltonian is effectively turned off when the pulses are applied to the system. Under this condition, the control-strength $c$ has to approach infinity when $\de$ goes to zero, which gives rise to both inconsistency in theory~\cite{JingPRA13} and inaccessibility in experiment. The underlying mechanism of BB control has been partially justified by the nonperturbative control~\cite{JWPRL15} when the integral of control pulses over time domain is sufficient large to enhance the survival probability of the system under control. It is a solution to attain a high-frequent exponential function $e^{-iC(s)}$. In the line of using longitudinal control to cancel the transversal error caused by the flip-flop Hamiltonian $L$, it is straightforward to extend the above protocol into a multi-level one.

Different from the BB control, quantum Zeno effect takes on the projection strategy, instead of an ideal unitary transformation, interpolating the unitary evolution of the whole system. After each period $\de_j$ of free evolution, the system is projected upon $|A(\de_j)\ra\la A(\de_j)|$ canceling all the errors caused by the system-environment interaction at the cost of a nondeterministic postselection~\cite{Zeno}. Similar to BB control, this protocol also does {\em not} depend on the sequence arrangement of the projection but the projection frequency. In reality, a particular realization of $\de_j$ could be random, noisy and even chaos~\cite{JingPRA13,JingSR13,JingSR14,JWPRL15}.

\subsection{Adiabaticity induced by control}\label{AIbC}

Through maintaining the system along the target path of $|A(t)\ra$, our control-activated LFP elevates the condition on achieving adiabatic passage of the system, where the quantum channel is realized through the time-dependent quantum eigenstate. Now the {\em Universal quantum control protocol} is applied in the following way. We first construct a time-dependent Hamiltonian for a nondegenerate system $H(t)=\sum_nE_n(t)|E_n(t)\ra\la E_n(t)|$ and let $|A(t)\ra=|E_0(t)\ra$ in lab frame up (at most) to a geometrical phase, where $E_n(t)$ and $|E_n(t)\ra$ are instantaneous eigenvalues and eigenvectors of $H(t)$, respectively. To obtain the one-component equation of motion for the amplitude of $|A(t)\ra$, it is instructive to rewrite the Hamiltonian in the adiabatic frame by $\mau=\sum_ne^{i\theta_n(t)}|E_n(0)\ra\la E_n(t)|$ with $\theta_n(t)\equiv\int_0^tdsE_n(s)$. We thus have
\begin{eqnarray*}
&& \ti{H}(t)=\mau H(t)\mau^\da+i\dot{\mau}\mau^\da \\
&& =\sum_{n,m,k}e^{i\theta_n(t)}|E_n(0)\ra\la E_n(t)|\cdot E_m(t)|E_m(t)\ra\la E_m(t)|\cdot e^{-i\theta_k(t)}|E_k(t)\ra\la E_k(0)| \\ &+&i\sum_{m,n}\left[i\dot{\theta}_m(t)e^{i\theta_m(t)}|E_m(0)\ra\la E_m(t)|+e^{i\theta_m(t)}|E_m(0)\ra\la\dot{E}_m(t)|\right]\cdot e^{-i\theta_n(t)}|E_n(t)\ra\la E_n(0)| \\
&=&\sum_nE_n(0)|E_n(0)\ra\la E_n(0)|-\sum_nE_n(0)|E_n(0)\ra\la E_n(0)|\\ &&+i\sum_{m,n}e^{i\theta_m(t)-i\theta_n(t)}\la\dot{E}_m(t)|E_n(t)\ra|E_m(0)\ra\la E_n(0)|\\
&=&-i\sum_n\la E_n(t)|\dot{E}_n(t)\ra|E_n(0)\ra\la E_n(0)|\\ && -i\sum_{m\neq n}e^{i\theta_m(t)-i\theta_n(t)}\frac{\la E_m(t)|\dot{H}(t)|E_n(t)\ra}{E_n(t)-E_m(t)}|E_m(0)\ra\la E_n(0)|.
\end{eqnarray*}
So that in language of Eq.~(\ref{Mt}), $\mathcal{X}(t)=[\psi_0(t),\psi_1(t),\psi_2(t),\cdots]'$, and
\begin{equation}
\mathcal{M}_{m\neq n}=-e^{-i[\theta_n(t)-\theta_m(t)]}\frac{\la E_m(t)|\dot{H}(t)|E_n(t)\ra}{E_n(t)-E_m(t)}.
\end{equation}
In particular, $P(t)=\psi_0(t)$ and $h(t)=-\la E_0(t)|\dot{E}_0(t)\ra$. In the case of unitary transformation into the adiabatic frame, the rotating LEO in Eq.~(\ref{rotLEO}) reads $\ti{R}_L=c(t)[2|E_0(0)\ra\la E_0(0)|-\mathcal{I}]$ and one can inversely derive the LEO in the lab frame as
\begin{eqnarray*}
&&R_L=\mathcal{U}^\da\ti{R}_L\mathcal{U} \\
&=&  c(t)\left[2\sum_{m,n}e^{-i\theta_m(t)}|E_m(t)\ra\la E_m(0)|\cdot|E_0(0)\ra\la E_0(0)|\cdot e^{i\theta_n(t)}|E_n(0)\ra\la E_n(t)|-\mathcal{I}\right] \\ &=& c(t)\left[2|E_0(t)\ra\la E_0(t)|-\mathcal{I}\right] \\
&=&c(t)\left\{\left[2\sum_{m,n}\la E_m(0)|E_0(t)\ra\la E_0(t)|E_n(0)\ra|E_m(0)\ra\la E_n(0)|\right]-\mathcal{I}\right\}.
\end{eqnarray*}
Applying LEO into controlling the system, the traditional condition for adiabaticity $|\la E_m|\dot{E}_n\ra|\ll|E_n-E_m|$ could be generalized to the vanishing accumulation of the product of $e^{-iC(s)}$ and $g(t,s)p(s)$ in Eq.~(\ref{gts}). With sufficiently fast-oscillating exponential function $e^{-iC(s)}$ via manipulating $h(t)\Rightarrow h(t)+c(t)$, i.e.,  $\mathcal{M}\Rightarrow\mathcal{M}+(-i\ti{R}_L)$, the adiabatic passage could be realized by rescaling the energy difference between the target state $|A(t)\ra$ and the other eigenstates of the system over time, and this control does {\em not} necessary boost this energy gap on time average. Our theorem is thus consistent with the original formalism of the adiabatic theorem and relaxes the slowly-varying condition. Additionally, it avoids the practical difficulty in some accelerated adiabatic passage, such as transitionless quantum driving~\cite{BerryJPA09} that requires to add a counter-adiabatic term into the original Hamiltonian (see also~\cite{Wang18} for the LEO in an experimental framework).

Beyond the adiabatic passage, any desired time-dependent LFP could be transformed into a time-independent one in the rotating frame after performing a proper unitary transformation. As long as the manipulation is highly frequent, there are unlimited numbers of strategies through which the LFP can be achieved since {\em the control is fully determined by $e^{-iC(t)}$, rather than $C(t)\equiv i\int_0^tdsh(s)$ (a large amplitude of this integral surely supports a high-frequent $e^{-iC(t)}$, but clearly it does not exhaust all the solutions) and even the details of the shape and arrangement of these pulses $c(t)$ presented in $h(t)$ under control.} For example, if $D(t)=a(t)|W\ra\la W|+\sum_jb_j(t)|W^\perp_j\ra\la W^\perp_j|$ [assuming both $W$ and $R$ in Eq.~(\ref{Mt}) are time-independent], where $\la W|W^\perp_j\ra=0$, i.e., $W$ is an eigenstate of $D$ and the eigenvalue is $a(t)$. Then in the case when $h'\equiv h-a$ is a constant (pure imaginary) number, it is found
\begin{equation}\label{DWaW}
p(t)=\frac{-h'+\De}{2\De}e^{\frac{-h'-\De}{2}t}+\frac{h'+\De}{2\De}e^{\frac{-h'+\De}{2}t},
\end{equation}
where $\De\equiv\sqrt{h'^2+4\la R|W\ra^2}$. When $|h'|$ could be so tuned that $|h'|t=2k\pi$, $k$ is an integer, the two oscillation frequencies in the expression of $p(t)$ will become sufficiently close to each other. Then $|p(t)|$ could be maintained as unit and a LFP emerges. It is consistent with and extends the previous result that $h(t)$ with a sufficient large magnitude~\cite{JWPRL15} overwhelming $2|\la R|W\ra|$ will suppress the decoherence.

\section{Applications}\label{app}

\subsection{A two-level system in accelerated adiabatic passage}

Following the conventions given by Sec.~\ref{AIbC}, a two-level-system state can be written as $|\psi(t)\ra=\psi_0(t)e^{-i\theta_0(t)}|E_0(t)\ra+\psi_1(t)e^{-i\theta_1(t)}|E_1(t)\ra$ in lab frame, where $|E_n(t)\rangle$'s, $n=0,1$, are instantaneous eigenstates. Choosing $|A(t)\ra=|E_0(t)\ra$ and suppose $E_1=-E_0=E/2$ without loss of generality, we have
\begin{equation}
\mathcal{M}=\left(\begin{array}{c|c}
      h=-\la E_0|\dot{E}_0\ra & R=-\la E_0|\dot{E}_1\ra e^{i\theta} \\ \hline
      W=-\la E_1|\dot{E}_0\ra e^{-i\theta} & D=-\la E_1|\dot{E}_1\ra
    \end{array}\right)
\end{equation}
in the adiabatic frame, where $\theta(t)\equiv\int_0^tdsE(s)=\theta_1(t)-\theta_0(t)$. Setting $p(t)=\psi_0(t)B(t)$ with $B(t)\equiv e^{\int_0^tds\la E_0(s)|\dot{E}_0(s)\ra}$, it yields a one-component dynamical equation as Eq.~(\ref{gts}):
\begin{equation}\label{gts2l}
\pa_tp(t)=e^{B(t)}\int_0^tdse^{-B(s)}g(t,s)p(s)=\int_0^tdsg'(t,s)p(s),
\end{equation}
where
\begin{eqnarray*}
g'(t,s)&=&\la E_0(t)|\dot{E}_1(t)\ra\la E_1(s)|\dot{E}_0(s)\ra\\&\times&\exp\left[\int_s^t ds^\prime \left(iE(s^\prime)+\la E_0(s')|\dot{E}_0(s')\ra-\la E_1(s')|\dot{E}_1(s')\ra\right)\right].
\end{eqnarray*}
For the control $H(t)\rightarrow[1+c(t)]H(t)$, in which the LEO reads
\begin{equation}
\ti{R}_L(t)=\frac{c(t)E(t)}{2}\left(\begin{array}{cc}
      -1 & 0 \\
      0 & 1
    \end{array}\right)=c(t)H(t).
\end{equation}
It will change the eigenvalues but not the eigenvectors. Once the frequency of the exponential function $e^{i\int_s^tds'E(s')}$ is effectively enhanced by $c(t)$, the integral in Eq.~(\ref{gts2l}) could vanish and then this two-level system would walk in an accelerated adiabatic path.

\subsection{A spin-spin-bath model}

Consider an electron spin coupled to a nuclear spin-$1/2$ bath through hyperfine interaction~\cite{hyperfine}:
\begin{equation}
H_{\rm tot}=\Om(t)S^z+\sum_n\om_nI_n^z+\sum_{n\alpha}J_n^{\alpha}(t)S^{\alpha}I_n^{\alpha}
+\sum_{nm\alpha}B_{nm}^{\alpha}I_n^{\alpha}I_m^{\alpha},
\end{equation}
where $\alpha=x,y,z$. In the single-exciton subspace and in the absence of the last term representing the inner coupling between the nuclear spins, we have
\begin{eqnarray*}
&& \mathcal{M}=-iH_{\rm tot} \\
&=&-i\left(\begin{array}{c|ccc}
\Om(t)-\sum_n\frac{J_n^z(t)}{2} & J_1^{\bot}(t) & J_2^{\bot}(t) & \cdots \\ \hline J_1^{\bot}(t) & \om_1-\frac{J_1^z(t)}{2} & 0 & 0 \\
J_2^{\bot}(t) & 0 & \om_2-\frac{J_2^z(t)}{2} & 0 \\
\cdots & \cdots & \cdots & \cdots \end{array}\right),
\end{eqnarray*}
where $J_n^\bot(t)\equiv J_n^x(t)+J_n^y(t)$. Now the $D$ matrix presents in the diagonal form. So that in the framework of our one-component quantum mechanics [see P-Q partitioning in Eq.~(\ref{Mt}) and the dynamical equation~(\ref{gts})], when the target state is chosen as the electron in the excited state and the nuclear spins in the ground state $|100\cdots\ra$, $g(t,s)=\sum_nJ_n^\bot(t)J_n^\bot(s)e^{-i\om_n(t-s)-\int_s^tds'J_n^z(s')/2}$, and $C(s)=i\int_0^sds'[\Om(s')-\sum_nJ_n^z(s')/2]$. One can then manipulate $\Om(t)$ to hold the population in the $P$-subspace. The control efficiency is relevant to the longitudinal Overhauser field $J_n^z(t)$~\cite{spinde}. In the presence of the inner coupling terms in nuclear spin bath, $h$, $W$ and $R$ do not vary while the diagonal terms of $D$ become $D_{nn}=\om_n-\frac{J_n^z(t)}{2}-\sum_{m\neq n}\frac{B_{nm}^z}{2}$ and the off-diagonal terms become $D_{nm}=B_{nm}^x+B_{nm}^y$. In this case, a protocol is to adjust the transversal Overhauser field (flip-flop term) $J_n^\bot(t)$ until $D|W\ra=a|W\ra$. Then one would achieve a similar result as Eq.~(\ref{DWaW}), so the electron spin is ensured to be maintained as the target state. It means that partial control on both system and bath promises to defeat decoherence.

\subsection{A multi-level system under non-Markovian environment}

For an $n$-level atom coupled to a non-Markovian bosonic bath whose correlation function is $\beta(t,s)$, we consider a genuine multilevel atomic system $H_{\rm sys}=\sum_{j=0}^{n-1}E_j(t)|j\ra\la j|$ and the system coupling operator, $S=\sum_{j=0}^{n-1}\ka_j|j\ra\la0|$. Using the non-Markovian quantum trajectory method~\cite{DiosiPLA97,DiosiPRA98,StrunzPRL99}, one can get an exact quantum-state-diffusion equation for this model:
\begin{equation}
\pa_t\psi_t(z^*)=\mathcal{M}\psi_t(z^*), \quad \mathcal{M}=-iH_{\rm sys}+Sz_t^*-S^\da\bar{O}(t),
\end{equation}
where $\bar{O}(t)=\sum_{j=1}^{n-1}F_j(t)|j\ra\la0|$. Coefficient functions $F_j(t)\equiv\int_0^tds\beta(t,s)f_j(t,s)$ satisfy $f_j(s,s)=\ka_j$ and $\pa_tf_j(t,s)=[i(E_0-E_j)+\sum_{k=1}^{n-1}\ka_k^*F_k(t)]f_j(t,s)$. The target state can be arbitrarily chosen as $|A\ra=\sum_{j=0}^{n-1}a_j|j\ra$, where $\sum_{j=0}^{n-1}|a_j|^2=1$. The fidelity $\mathcal{F}\equiv\la A|\rho|A\ra=M[\la A|\psi(z^*)\ra\la\psi(z^*)|A\ra]$, where $M[\cdot]$ means ensemble average, measuring the control efficiency, is then found to depend on the population rather than the amplitude of the initial state:
\begin{eqnarray*}
\mathcal{F}(t)&=&|a_0|^4e^{-\sum_{j=1}^{n-1}[\bar{F}_j(t)+\bar{F}^*_j(t)]}+\sum_{j,k\geq1,k\neq j}|a_j|^2|a_k|^2\\ &+&
|a_0|^2\sum_{k=1}^{n-1}|a_k|^2\left[e^{-\sum_{j=1}^{n-1}\bar{F}_j(t)}+e^{-\sum_{j=1}^{n-1}\bar{F}^*_j(t)}\right] \\
&+&\sum_{j=1}^{n-1}|a_j|^2\left\{|a_j|^2+\int_0^tds|a_0|^2[F_j(s)+F_j^*(s)]
e^{-\sum_{j=1}^{n-1}[\bar{F}_j(s)+\bar{F}^*_j(s)]}\right\},
\end{eqnarray*}
where $\bar{F}_j(t)\equiv\int_0^tdsF_j(s)$. It implies that the control fidelity depends only on the populations.

\begin{figure}[htbp]
\centering
  \subfigure{\label{g2}\includegraphics[width=2.7in]{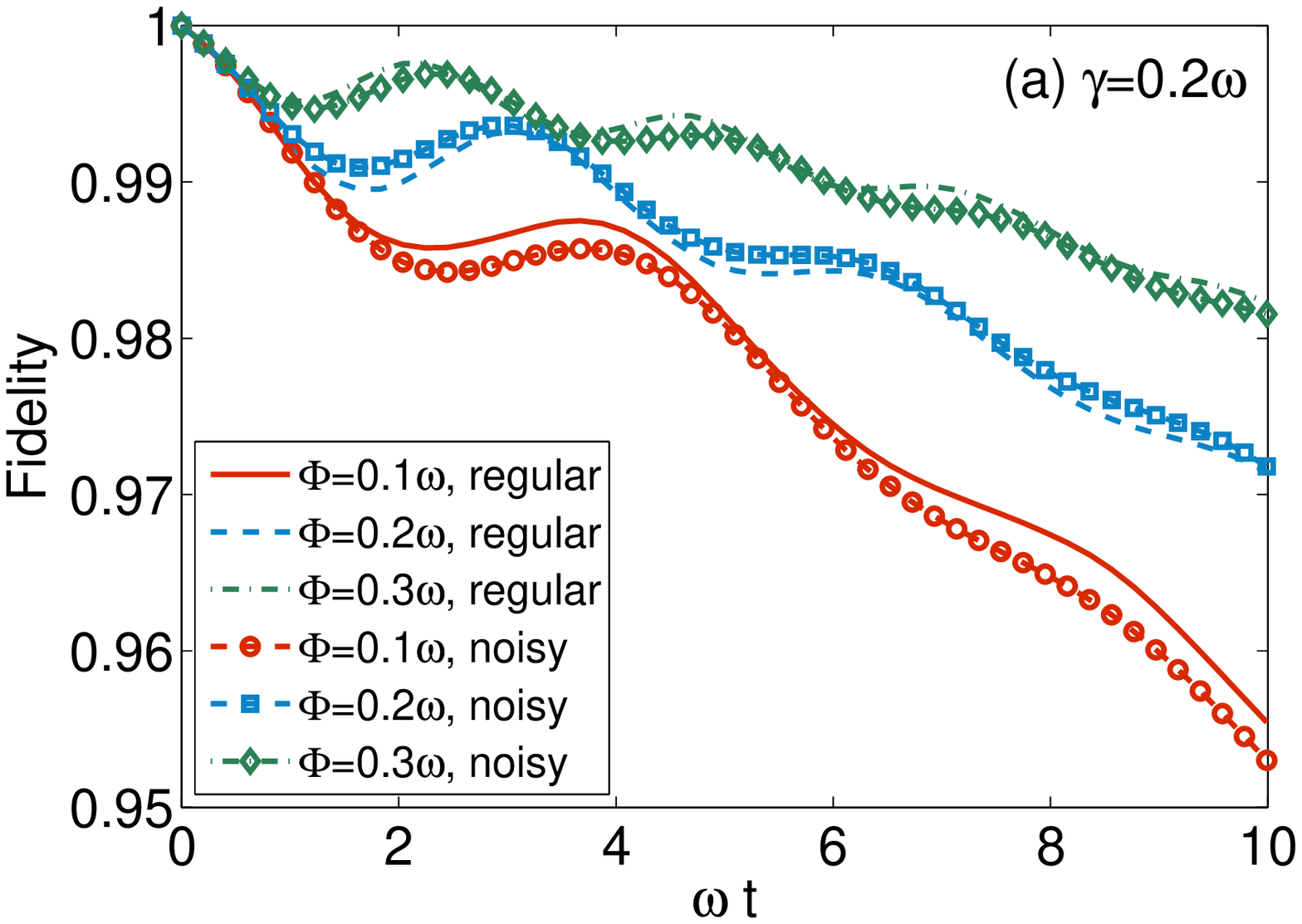}}
  \subfigure{\label{g5}\includegraphics[width=2.7in]{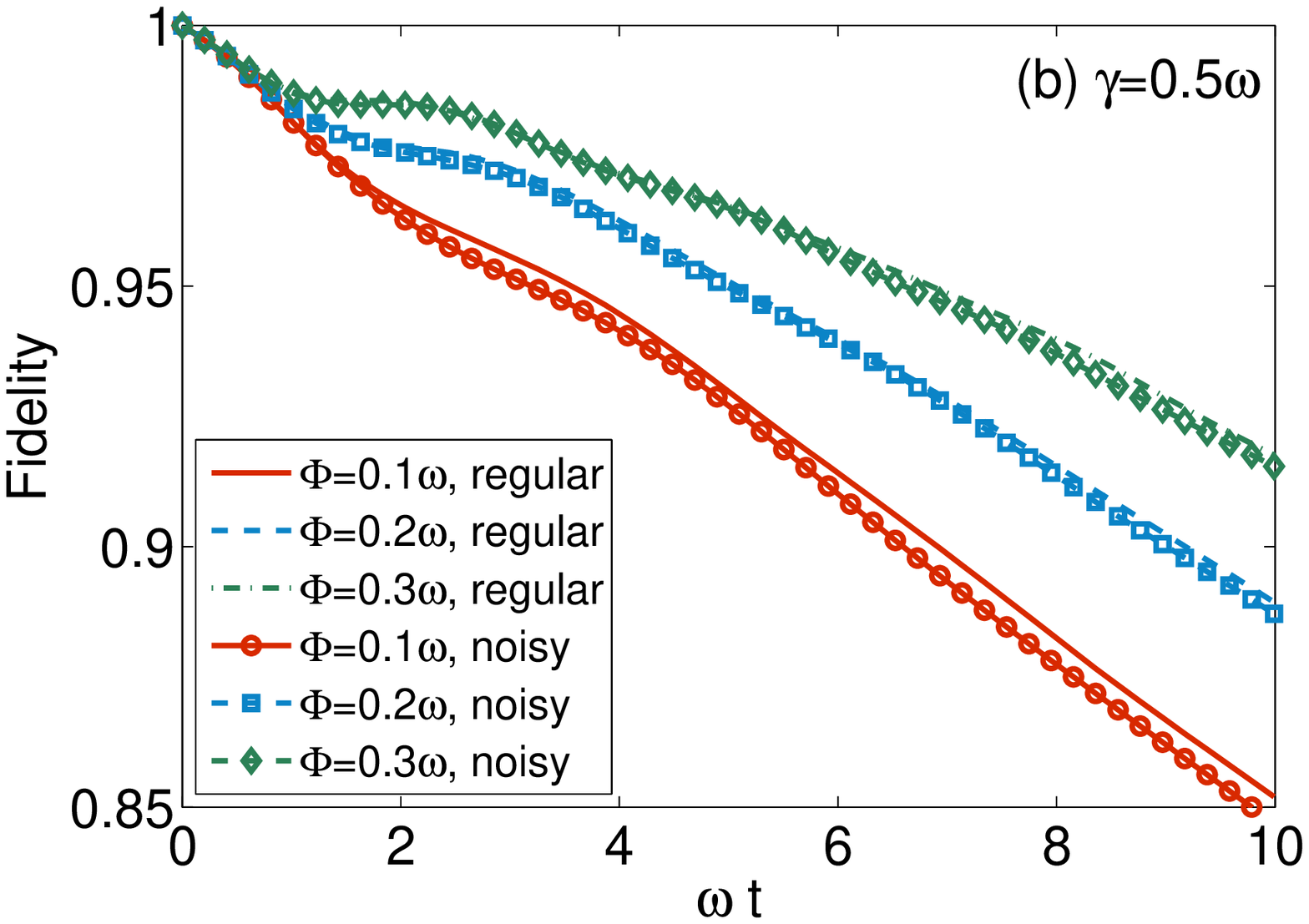}}
\caption{Fidelity dynamics of a $10$-level atomic system under control of regular and noisy pulse sequences. The target state $|A\ra=\sum_{j=0}^{n-1}a_j|j\ra$ is such chosen as $|a_j|^2=1/10$. The parameters are chosen as $\ka_j=0.1\om$, $E_{j\neq0}=0$ and $E_0(t)=\om$. For regular rectangular pulse, $|c(t)|=\Phi/\De$ for $m\tau-\De\leq t\leq m\tau$, where $m\geq1$ is an integer; otherwise, $|c(t)|=0$. $\Phi$, $\De$, and $\tau$ are the strength, duration and period of pulse, respectively. Here $\tau=0.02\om t$ and $\De/\tau=0.5$. For noisy pulse, $|c(t)|\rightarrow|c(t)|[1+G\mathcal{N}(t)]$ where $G$ (here $G=50\%$) is a dimensionless parameter measuring the white noise strength $\mathcal{N}(t)\in(-1, 1)$. The environment correlation function is taken as $\beta(t,s)=\frac{\ga}{2}e^{-\ga|t-s|}$, where $\ga$ is inversely proportional to the environmental memory time. A smaller $\ga$ indicates a stronger non-Markovian environment.}
\end{figure}

In Figs.~\ref{g2} and \ref{g5}, we plot the fidelity of a multilevel system under control of regular and noisy sequences of pulse in environments with different environmental memory parameters. It is shown that the fidelity is enhanced with increasing pulse strength $\Phi$, which is linearly proportional to the absolute value of control integral $C(t)$. Note the average of control integral is kept vanishing for the sign of $c(t)$ is switched periodically (see the lines with no markers) or randomly (see the lines with markers). One can find that the effect of regular control is nearly the same as the noisy sequence, which does not change $e^{-iC(t)}$ in terms of ensemble average and long time simulation. Comparing Figs.~\ref{g2} and \ref{g5}, it is reliable to estimate that any control will gradually lose its effect when the environment becomes more and more memoryless (note $\ga\rightarrow\infty$ indicates a Markovian environment). The LFP of Figs.~\ref{g2} and \ref{g5} is represented by $|a_j|^2=1/n$, while for an arbitrary target state, $\mathcal{M}$ in our control protocol can be found by a corresponding rotating as $\mathcal{U}|A\ra=|\ti{A}\ra$. 

\section{Conclusion}\label{conc}

In this work, we set up a general framework that allows one to follow and control a quantum system, open or closed, along a desired leakage-free path, presented by a one-dimensional dynamical equation addressing the time-dependent target state. A common mechanism subtly underlying existing quantum control protocols, including BB control, Zeno effect and adiabatic passage, is brought to light as a general condition for dynamical leakage-free paths. As long as the exponential function of a phase provided by the integral or cumulation over the control pulse is featured with a sufficient large frequency, the LFP can be realized by leakage elimination operation. As an active protocol that is not confined by the structure of the total Hamiltonian, it would be versatile to accommodate arbitrary linear non-Markovian equations of motion for open quantum systems. Moreover, upon proper control over system Hamiltonian or partial control over system-environment interaction Hamiltonian, the target state $|A\ra$ or $|A\ra\ra$ could be extended into a more general LFP in multi-dimensional space that is able to perform more quantum processing, such as the non-Abelian geometric quantum gate in degenerate subspace, the shortcut to quantum state transmission~\cite{Wang16}, the speeding up holonomic quantum computation in decoherence-free subspace~\cite{Pavel16}, the quantum search algorithm and almost-exact state transfer in non-Markovian environments~\cite{Ren20,Wang20,M20}. 

\section*{Acknowledgments}

We acknowledge grant support from the National Science Foundation of China (Grants No. 11974311 and No. U1801661), the Basque Government IT986-16, Spanish Government PGC2018-095113-B-I00 (MCIU/AEI/FEDER, UE).

\end{document}